\documentclass[acmsmall, nonacm]{acmart}
\usepackage{amsmath}
\usepackage[parfill]{parskip}

\AtBeginDocument{%
  \providecommand\BibTeX{{%
    \normalfont B\kern-0.5em{\scshape i\kern-0.25em b}\kern-0.8em\TeX}}}

\setcopyright{acmcopyright}
\copyrightyear{2024}

\begin{document}

\title{A Comprehensive Review: Applicability of Deep Neural Networks in Business Decision Making and Market Prediction Investment}

\author{Viet Trinh}
\email{tqviet@uel.edu.vn}
\affiliation{%
  \institution{University of Economics and Law}
  \streetaddress{Vietnam National University HCM}
  \city{Ho Chi Minh}
  \country{Vietnam}
}

\renewcommand{\shortauthors}{Viet Trinh}

\begin{abstract}
Big data, both in its structured and unstructured formats, have brought in unforeseen challenges in economics and business. How to organize, classify, and then analyze such data to obtain meaningful insights are the ever-going research topics for business leaders and academic researchers. This paper studies recent applications of deep neural networks in decision making in economical business and investment; especially in risk management, portfolio optimization, and algorithmic trading. Set aside limitation in data privacy and cross-market analysis, the article establishes that deep neural networks have performed remarkably in financial classification and prediction. Moreover, the study suggests that by compositing multiple neural networks, spanning different data type modalities, a more robust, efficient, and scalable financial prediction framework can be constructed.
\end{abstract}


\ccsdesc[500]{Information systems~Data management systems}
\ccsdesc[500]{Information systems~Information systems applications}
\ccsdesc[300]{Information systems applications~Decision support systems}

\keywords{deep learning, neural networks, algorithmic trading, risk management, fraud detection}

\maketitle

\section{Introduction}
Vast volume of data in daily business operations, either in a structured (e.g., numerical values in reports) or an unstructured format (e.g., documents, emails, financial transactions), has posed a significant challenge in decision making. It is not the amount of data that is important, in fact, it is what insights such data imply that matters. In the past decades, technologies used in Artificial Intelligence, especially Machine Learning and Deep Learning, have been widely employed to improve efficiency in most aspects of life: from facial detection in image processing to text predictions in natural language processing. Recently, there is a growing interest in applying artificial intelligence into business administration, financial management, and economical policy making. 

The goal of this article is to convey a fundamental understanding of deep learning neural networks, and to discuss their utilization and limitation in economics, finance, and business. Our study contributes to the literature in the following ways. In the section \ref{sec:prelim-discussion}, we comprehensively and systematically review different deep learning paradigms, as well as introducing some popular neural networks. Section \ref{sec:financial-app} discusses several data collection sources, and various applications of deep learning models in financial prediction. Next, section \ref{sec:discussion} outlines limitation of financial data and current challenges in decision making in economical business and investment. Finally, section \ref{sec:conclusion} concludes our work with promising future research direction.

\section{Preliminary discussion on deep learning}
\label{sec:prelim-discussion}
Since we study the applicability of neural networks in business and investment in this paper, it is worth briefly mentioning current deep learning techniques and understanding several neural network models in use. Not to focusing heavily on technicality of such models, this section aims to introduce a neural network in a nutshell: what the input is, what the output is, and what the middle layers of a neural network are responsible for. We will conclude the section with our thoughts on simultaneously employing several models in a join-effort manner in hope for more accurate outcomes.

\subsection{Learning paradigms}
\begin{figure}[h]
  \centering
  \includegraphics[width=\linewidth]{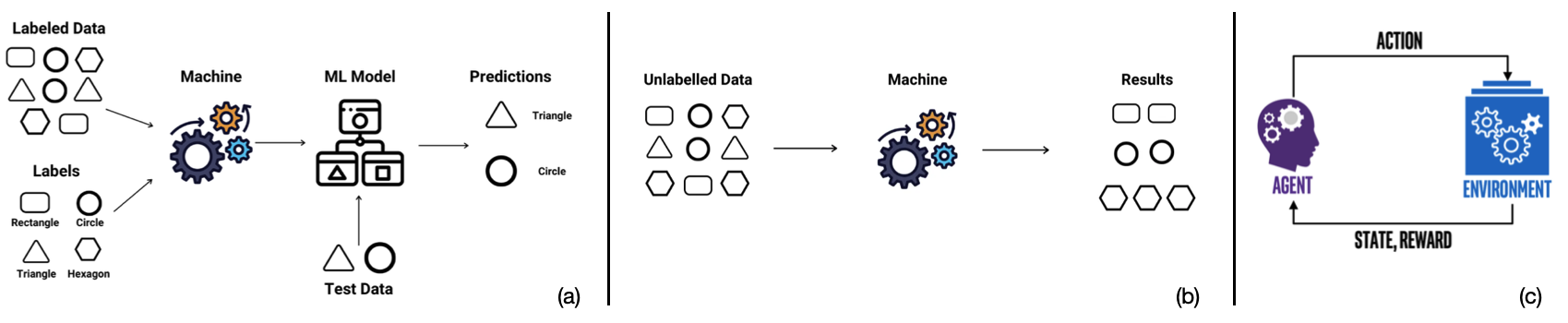}
  \caption{\label{fig:three-paradigms}Three paradigms in deep learning: supervised, unsupervised, and reinforcement. (a) In supervised learning, models are trained on labeled data to mostly perform the task of classification (where results are categories) and regression (where results are continuous values); (b) in unsupervised learning, models are trained on unlabeled data to identify associated relationships in a dataset, insightful cluster data group, and reduce high dimensional data form.}
\end{figure}

Nowadays, learning algorithms are mostly considered as constructing and employing deep neural networks which is pre-trained on a very large data set, and is then fine-tuned to follow instructions (e.g: define a policy or an optimal path) or to solve specific tasks (e.g: object detection and classification). Based on the way algorithms learn from data, it can be divided into three learning paradigms: \textbf{supervised learning}, \textbf{unsupervised learning}, and \textbf{reinforcement learning}. The primary differences among those three are in the types of result being produced, and the way neural network models validate and calibrate those results. Figure \ref{fig:three-paradigms} illustrates such differences of those learning paradigms.

\begin{itemize}
	\item \textbf{Supervised learning} involves training a model from labeled data. In other words, a neural network learns to predict the future output based on what happened in the past input. It will identify any patterns or trends (also known as \textit{weight}) that are historically correlated with the ground-truth outcomes and then use them to perform future prediction. 
	\item \textbf{Unsupervised learning} analyzes and clusters unlabeled data sets; aiming to discover hidden patterns in data without the need for human intervention. This type of learning is mostly used for grouping data in a meaningful way, representing hight-dimensional data set in a low-dimensional form, and associating relationships between variables in a large dataset.
	\item \textbf{Reinforcement learning}, in the other hand, does not rely on any pre-learned pattern. Instead, it mimics the trial-and-error learning process that humans use to achieve its goals. Specifically, neural network models in reinforcement learning use a reward-and-punishment paradigm as they processes data. The algorithm constantly tries new things to learn and improving upon current approaches to maximize the defined reward.
\end{itemize}

\subsection{Neural network models}
A neural network is a nonlinear function represented by a collection of neurons. These collections are normally categorized into layers; connected by operators such as filters, poolings and gates. Its main objective is to map an input in domain $\mathbb{R}^n$ to an output in domain $\mathbb{R}^m$ ($n, m \in \mathbb{Z}^+$). Intermediate layers of neurons act as key features extraction for such mapping. Currently, the three most popular neural network architectures are Fully connected neural networks (FNN), Convolutional neural networks (CNN), and Recurrent neural networks (RNN).

\begin{figure}[h]
  \centering
  \includegraphics[width=\linewidth]{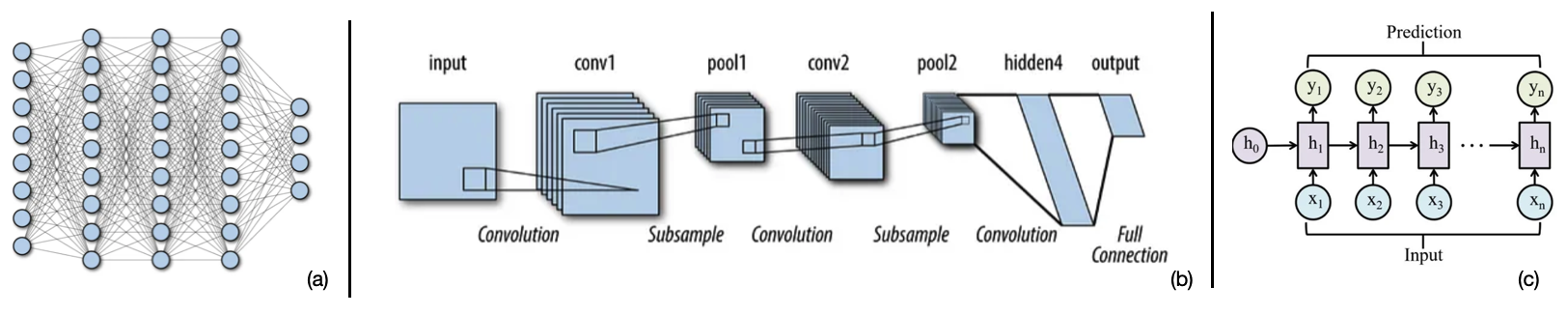}
  \caption{\label{fig:neural-networks}Neural network models: (a) Fully connected neural network; (b) Convolutional neural network;  (c) Recurrent neural network (RNN).}
\end{figure}

FNN, shown in figure \ref{fig:neural-networks}(a), is the simplest architecture in which all neurons of the previous layer are connected to all neurons of the next layer \cite{chen2021neural}. The major advantage of FNN is that there are no special assumptions needed to be made about the input. This is also the disadvantage of FNN since it tends to have weaker performance for a specific task. CNN (figure \ref{fig:neural-networks}(b)) makes an explicit assumption that the inputs are images, allowing encoding certain properties into the model architecture and serving the specific task of classification and regression \cite{jiang2023re}. In RNN, layers of neurons form a directed cycle, instead of a straight model like FNN or CNN. This makes RNN to be widely used in processing sequential data, time series data, and arbitrary inputs (figure \ref{fig:neural-networks}(c)).  

In economics and finance settings, CNN and RNN are dominant network candidates employed for prediction task. Specifically, CNNs have been applied to credit risk, price, and financial transactions prediction \cite{jiang2023re, mukherjee2023stock, chen2021neural}. Feature set inputs are mostly images containing candlestick charts or price and trading volume graphs. Similarly, RNNs have been used in classifying financial stress, managing fraud transactions, and trading prediction. Unlike CNNs where input sets are images, RNNs take in historical series data of price, volume, annual return, index data, etc in a numerical format \cite{sarkar2021lstm, de2021investment, jiang2021applications}. Current research has suggested to stack FNNs on top of CNNs and RNNs for a better feature extraction, noise and nullable cancellation of missing data, and voting mechanism for what to feed-in in the next training iteration \cite{jing2021hybrid, carta2021multi}.

\section{Financial applications}
\label{sec:financial-app}
It is inarguable that a vast amount of information and its accessibility have posed a significant concern in the field of economics and business. Data is constantly and simultaneously generated from different sources: financial transactions, inventory log, customer feedbacks, social network news, etc. How to organize, classify, and analyze such data to obtain meaningful insights, helping in making a right strategical decision, is the ever-going challenge for business leaders and academic researchers.

In the study of \cite{singh2022reinforcement}, it has been found that there are 460 articles published from 2014 to 2021, suggesting different approaches in machine learning and deep learning to tackle multiple domains of problems in economics and business. Specifically, 260 out of 460 such articles focus on the category of risk management, portfolio optimization and trading and assets. Data are collected in multiple forms from different sources:
\begin{itemize}
	\item \textbf{market data} include trading activities in stock market (e.g: stock prices, stock volume, transactions)
	\item \textbf{economic data} consist of macroeconomics circumstances of regions or countries (e.g: Consumer Price Index, Gross Domestic Product)
	\item \textbf{image data} contain not only knowledge graph representing companies's performance (e.g: revenue dashboard), but also candlestick charts of stock and gold fluctuation
	\item \textbf{analytics data} extract from task-specific reports such as risk analysis, credit score review, or property appraisal
	\item \textbf{text data} refer to text generated by end users on new, web search, social media posts, etc. Such data contain a strong sentiment information (e.g: positive, negative, neutral) and is a significant factor in prediction
\end{itemize}

Table \ref{tab:apply-ai} lists current interests in utilizing deep learning models for insightful decision making in economics, finance, and business. In this section, we discuss an overview of various applications of deep learning models in financial prediction, and what type of input could be useful in training and fine-tuning a model.

\begin{table}
\small
  \caption{Overview of studies that utilizes deep learning models in economics, finance, and business}
  \label{tab:apply-ai}
  \begin{tabular}{ll}
    \toprule
    Category&Objective\\
    \midrule
    Credit risk& General consumer default\\
    & Credit card delinquency\\
    & Bill payment patterns\\
    & Fintech loan default\\
    & Mortgage loan risk\\
    Fraud& Accounting fraud from financial statements\\
    & Accounting fraud from annual reports\\
    & Acquisitions\\
    Trading & Buy and sell decisions\\
    & Sentiment analysis\\
    & Fund performance\\
    & Portfolio optimization\\
    & Exchange rates direction of changes\\
  \bottomrule
\end{tabular}
\end{table}

\subsection{Risk management}
Risk management, consisting of \textit{Risk Assessment} and \textit{Fraud Detection}, has lately gained a great interest in both industrial and academic research communities. Risk assessment studies credit scoring, financial stress, loan application, mortgage lending decision, and business evaluation. Fraud detection aims to identify faulty transactions, credit card exploitation, money laundering, tax evasion, and insurance claim fraud. The end goal of these studies is to help determining the probability of a risk to be occurred. While it is not guaranteed that a predicted risk is a true risk, neural network models using in these research can significantly providing insights to decision makers. 

For example, in credit score classification, Zhu et al. \cite{zhu2018hybrid}, transformed 2-D consumer expense data into a 2-D pixel matrix, and then used the resulting images as dataset for training and validating a CNN-based model. In another work, \cite{jing2021hybrid} classified a bank risk by employing a LSTM model to explore word sequences extracted from financial reports. Furthermore, there are several approaches in combing FNN and LSTM models to analyze character sequences in financial transactions and to detect anomaly in consumer spending behavior, aiming to detect fraudulent activities \cite{sarkar2021lstm, zohuri2023artificial, singh2022reinforcement}. 

\subsection{Portfolio optimization}
In portfolio optimization, traders need to analyze, select, and trade the currently best financial assets to maximize benefits, as well as minimizing any potential risk. One of the most promising works in this area is the multi-layer and multi-ensemble stock trader neural network, introduced by Carta and his group in \cite{carta2021multi}. His method starts by pre-processing market data with hundreds of deep neural networks, generates potential stock signals with a reward-based classifier, and then concludes the trading decision through a voting system. Such reward-based classifier, also known as meta-learner, leads to better trading results and less overfitting during training and evaluation. 

In a similar work, Lee has defined a threshold-based portfolio for traders to diversify investments by employing policy-based algorithms in reinforcement learning \cite{lee2020threshold}. In such learning, the state variables are past asset returns, current asset prices, current asset inventory, and remaining balance. The control factors are mainly proportion of wealth invested in each portfolio. 

\subsection{Algorithmic trading}
Algorithmic trading (AT) has been considered a hugely successful application in the recent years, due to its forecasting models for trigger buy-sell signals. Specifically, if the target objective is to predict the specific price values, AT can be treated as regression problem. Whereas, AT can be considered as a classification problem if the end goal is to predict the market price movement direction. Most studies in AT are performed for daily prediction instead of intraday prediction (e.g., 5-minute or hourly prediction) \cite{thakkar2021comprehensive}. 

Recently, Li et al. \cite{li2021day} proposed a framework, combining both CNN and LSTM models, to forecast stock prices. In this study, CNN architecture was used for stock selection, and LSTM was employed for price prediction. In other work, social media knowledge is used in conjunction with market data to establish a framework for analyzing and predicting trends in stock prices \cite{mehta2021harvesting}. The motive of this study is to provide accurate predictions of the stock market values for consecutive days. In a different approach of predicting the near future market stock prices, \cite{janiesch2021machine} focused on generating synthetic images in 2‐D histograms from market data spanning over a few days, and then used CNN model to predict the time‐series stock value of the next following day.

\section{Discussion}
\label{sec:discussion}
Our review establishes that employing deep neural networks has yielded a remarkable result in economical and financial prediction purposes. The challenges of categorizing and decision making on a huge amount of financial data can be easily reduced into the problems of detection, classification, and regression in deep learning. Depending on the source of collected data (e.g., market data, economic data, image data, analytics, data, and text data), different neural network models (e.g., FNN, CNN, RNN-LSTM) yield different accurate predicted outcomes. For example, for time-series stock prices, LSTM is proven to be the best choice for daily stock values prediction. Whereas, for candlestick charts or transaction graphs, CNN is a better fit for training and evaluating. 

Additionally, it is concludable that different feature section methods will lead to different prediction outcomes, as deep neural networks heavily rely on key features for mapping between input and output. Beside standalone networks such as FNN, CNN, or LSTM, it is also proven that stacking one on top of another, or using multiple types of neural networks in parallel, can lead to better trading results and less overfitting when exploring training parameters. Also, several research works have suggested that predicting stock market movements should include social networking knowledge as a configurable input since it reflects perfectly the public's views of a current market. Recent advancements in deep learning framework can provide a more accurate and robust approach to handle stock market predictions, based on both social feedbacks and historical time-series stock prices.
                          
\subsection{Challenges}
Although there are advancements in applying trainable neural networks to improve risk detection, asset optimization, and stock trading, there also exists limitation. Specifically, these frameworks are trained on financial data that contain sensitive, private, and personal information. As a result, predicted outputs may compromise privacy, and hence, leading to limit accessibility to future training dataset. Another challenge comes from availability and integrity of real-time data. We all know that training or fine-tuning a network takes time (in a matter of days), however market data fluctuate in the unit of hours or minutes. This implies that a successfully established model for a stock prediction, in some sense, might be trained on out-dated dataset comparing to the real world situation. Also, recent works have shown that a different specific stock market is considered in each different study. There is no cross-stock market training or evaluating research. This is due to stock markets differ from each other because of the trading rules and policies. Most of existing studies in risk management and trading suggestion focus on employing RNN/LSTM architectures for time-series market data. This is understandable in a sense that stock market prices come in sequential order over a time period. There is not a lot of opportunities for training or fine-tuning other well-known neural network types due to limitation on input data type.

\subsection{Future research direction}
In our opinion, multiple neural networks or composition of neural networks are not fully studied for the applicability of artificial intelligence in economics and finance. Generally, most of the problems in risk management, portfolio optimization, and tradings can be broken down into two steps: Data Processing and Prediction. If we are able to convert a single data type (e.g., credit scores, stock prices, asset values) into other data formats, we can think of a framework consisting of multiple deep neural networks that acts separately on corresponding data types, and then votes on generated outputs of these neural networks to determine the final prediction. For example, transformer networks are widely used in natural language processing for sentiment analysis of customer feedback, but is less discussed for financial news analysis.

We agree that stock markets differ among regions and countries because of trading policies, however, they may share some common phenomenon that can be leveraged for prediction by approaches such as transfer learning. Future research work should look into cross-market analysis, either training on multiple stock markets or fine-tuning and transfer learning from one to another. Additionally, it is extremely beneficial to include social networking knowledge as meta-data input in forecasting stock market movement, as it is one of the main driven factors nowadays. Impact on the stock market from social networkings must not be neglected.

\section{Conclusion}
\label{sec:conclusion}
This paper reviews the current applicabilities of deep neural networks in decision making in business and investment, specially in three sub-domains: risk management, portfolio optimization, and algorithmic trading. Our work establishes that deep neural networks have made a significant improvement on predicted outcomes with high accuracy, comparing to traditional stochastic approaches. Set aside challenges in data privacy and availability for cross-market analysis, we believe that such networks allow us to develop a more robust and efficient framework to confidently determine the final prediction by reasoning different modalities of data types and voting on the common arising phenomenon. With the ever-growing of big data and deep learning platforms, accuracy on real-time prediction is the future research direction in investment and market tradings.

\bibliographystyle{ACM-Reference-Format}
\bibliography{bibliography}

\end{document}